\documentclass[twocolumn]{article}
\usepackage{sample}

\def\BibTeX{{\rm B\kern-.05em{\sc i\kern-.025em b}\kern-.08em
    T\kern-.1667em\lower.7ex\hbox{E}\kern-.125emX}}

\title{Random Indexing K-tree}
\author{
{\em Christopher M. De Vries \hspace{10px} Lance De Vine \hspace{10px} Shlomo Geva}\\[1ex]
Faculty of Science and Technology\\Queensland University of Technology\\Brisbane, Australia\\[1ex]
{\em chris@de-vries.id.au \hspace{10px} l.devine@qut.edu.au \hspace{10px} s.geva@qut.edu.au} 
}
\date{}

\usepackage{amsfonts}
\usepackage{amsmath}
\usepackage{graphicx}

\begin{document}

\maketitle
\thispagestyle{empty}

        \begin{figure}[b]
	~\\
        \noindent
        {\small\raggedright\bf
        Proceedings of the 14th Australasian 
	Document Computing Symposium,
	Sydney, Australia,
        4 December 2009.
	Copyright for this article remains with the authors.
        }
        \end{figure}

\paragraph*{Abstract}
\noindent
{\it 
Random Indexing (RI) K-tree is the combination of two algorithms for clustering. Many large scale problems exist in document clustering. RI K-tree scales well with large inputs due to its low complexity. It also exhibits features that are useful for managing a changing collection. Furthermore, it solves previous issues with sparse document vectors when using K-tree. The algorithms and data structures are defined, explained and motivated. Specific modifications to K-tree are made for use with RI. Experiments have been executed to measure quality. The results indicate that RI K-tree improves document cluster quality over the original K-tree algorithm.
} 

\paragraph*{Keywords} 
Random Indexing, K-tree, Dimensionality Reduction, B-tree, Search Tree, Clustering, Document Clustering, Vector Quantization, k-means

\section{Introduction}

The purpose of this paper is to present and analyse the combination of Random Indexing (RI) with the K-tree algorithm.  Both RI and K-tree adapt to changing data and decrease the cost of computationally intensive vector based applications. This combination is particularly suitable to the representation and clustering of very large document collections.  Documents are typically represented in vector space as very sparse high dimensional vectors. RI can reduce the dimensionality and sparsity of this representation.  In turn, the condensed representation is highly effective when working with K-tree. The paper is focused on determining the effectiveness of using RI with K-tree through experiments and comparative analysis of results.

Sections \ref{sec:ktree} to \ref{sec:results} discuss K-tree, Random Indexing, Document Representation, Experimental Setup and Experimental results respectively. The paper ends with a conclusion in Section \ref{sec:conclusion}.

\section{K-tree}
\label{sec:ktree}

K-tree \cite{DeVries2009,ktreeweb} is a height balanced cluster tree. It was first introduced in the context of signal processing by Geva \cite{Geva2000}. The algorithm is particularly suitable to clustering of large collections due to its low complexity. It is a hybrid of the B$^+$-tree and k-means algorithm. The B$^+$-tree algorithm is modified to work with multi dimensional vectors and k-means is used to perform node splits in the tree. K-tree is also related to Tree Structured Vector Quantization (TSVQ) \cite{Gersho1993}. TSVQ recursively splits the data set, in a top-down fashion, using k-means. TSVQ does not generally produce balanced trees.

K-tree achieves its efficiency through execution of the high cost k-means step over very small subsets of the data. The number of vectors clustered during any step in the K-tree algorithm is determined by the tree order (usually $\ll$ 1000) and it is independent of collection size. It is efficient in updating the collection while maintaining clustering properties through the use of a nearest neighbour search tree that directs new vectors to the appropriate leaf node.

The K-tree forms a hierarchy of clusters. This hierarchy supports multi-granular clustering where generalisation or specialisation is observed as the tree is traversed from a leaf towards the root or vice versa. The granularity of clusters can be decided at run-time by selecting clusters that meet criteria such as distortion or cluster size.

\subsection{K-tree and Document Clustering}

The K-tree algorithm is well suited to clustering large document collections due to its low time complexity. The time complexity of building K-tree is O(n log n) where n is the number of bytes of data to cluster. This is due to the divide and conquer properties inherent to the search tree. De Vries and Geva \cite{DeVries2008,DeVries2009} investigate the run-time performance and quality of K-tree by comparing results with other INEX submissions and CLUTO \cite{Karypis2002}. CLUTO is a popular clustering tool kit used in the information retrieval community. K-tree has been compared to k-means, including the CLUTO implementation, and provides comparable quality and a marked increase in run-time performance. However, K-tree forms a hierarchy of clusters and k-means does not. Comparison of the quality of the tree structure will be undertaken in further research. The run-time performance increase of K-tree is most noted when a large number of clusters are required. This is useful in terms of document clustering because there are a huge number of topics in a typical collection. The on-line and incremental nature of the algorithm is useful for managing changing document collections. Most clustering algorithms are one shot and must be re-run when new data arrives. K-tree adapts as new data arrives and has the low time complexity of O(log n) for insertion of a single document. Additionally, the tree structure also allows for efficient disk based implementations when the size of data sets exceeds that of main memory.

\subsection{K-tree Definition}

K-tree builds a nearest neighbour search tree over a set of real valued vectors $V$ in $d$ dimensional space.
\begin{equation}
\forall v \in V : v \in \mathbb{R}^d
\end{equation}
It is inspired by the B$^+$-tree where all data records are stored in leaf nodes. Tree nodes, $N$, consist of a sequence of (vector, child node) pairs of length $l$. The tree order, $m$, restricts the number of vectors stored in any node to between one and $m$.
\begin{equation}
1 \leq l \leq m
\end{equation}
\begin{equation}
N = \langle (v_1, c_1),...,(v_l,c_l) \rangle
\end{equation}
The tree consists of two types of nodes. Leaf nodes contain the data vectors that were inserted into the tree. Internal nodes contain clusters. A cluster vector is the mean of all data vectors contained in the leaves of all descendant nodes (i.e. the entire cluster sub-tree).  This follows the same recursive definition of a B$^+$-tree where each tree is made up of a set of smaller sub-trees. Upon construction of the tree, a nearest neighbour search tree is built in a bottom-up manner by splitting full nodes using k-means \cite{Lloyd1982} where $k = 2$. As the tree depth increases it forms a hierarchy of ``clusters of clusters'' from the root to the above-leaf level. The above-leaf level contains the finest granularity cluster vectors. Each leaf node stores the data vectors pointed to by the above-leaf level.  The efficiency of K-tree stems from the low complexity of the B$^+$-tree algorithm, combined with only ever executing k-means on a relatively small number of vectors, defined by the tree order, and by using a small value of $k$.

\subsection{Modifications to K-tree}
\label{sec:modifications}

The K-tree algorithm was modified for use with RI. This modified version will be referred to as ``Modified K-tree'' and the original K-tree will be referred to as ``Unmodified K-tree''.

All the document vectors created by RI are of unit length in the modified K-tree. Therefore, all centroids are normalised to unit length at all times. The k-means used for node splits in K-tree was changed to use randomised seeding and restart if it did not converge within six iterations. The process always converged quickly in our experiments; although it is possible to constrain the number of restarts we did not find this to be necessary.

The original K-tree algorithm does not modify any of the centroids. They are simply the means of the vectors they represent. The k-means implementation runs to complete convergence and seeds centroids via perturbation of the global mean. To create two seeds the global mean is calculated and then the two seeds are created by moving away from the mean in opposite directions.

\subsection{K-tree Example}

\begin{figure}
\centering
\includegraphics[width=75mm]{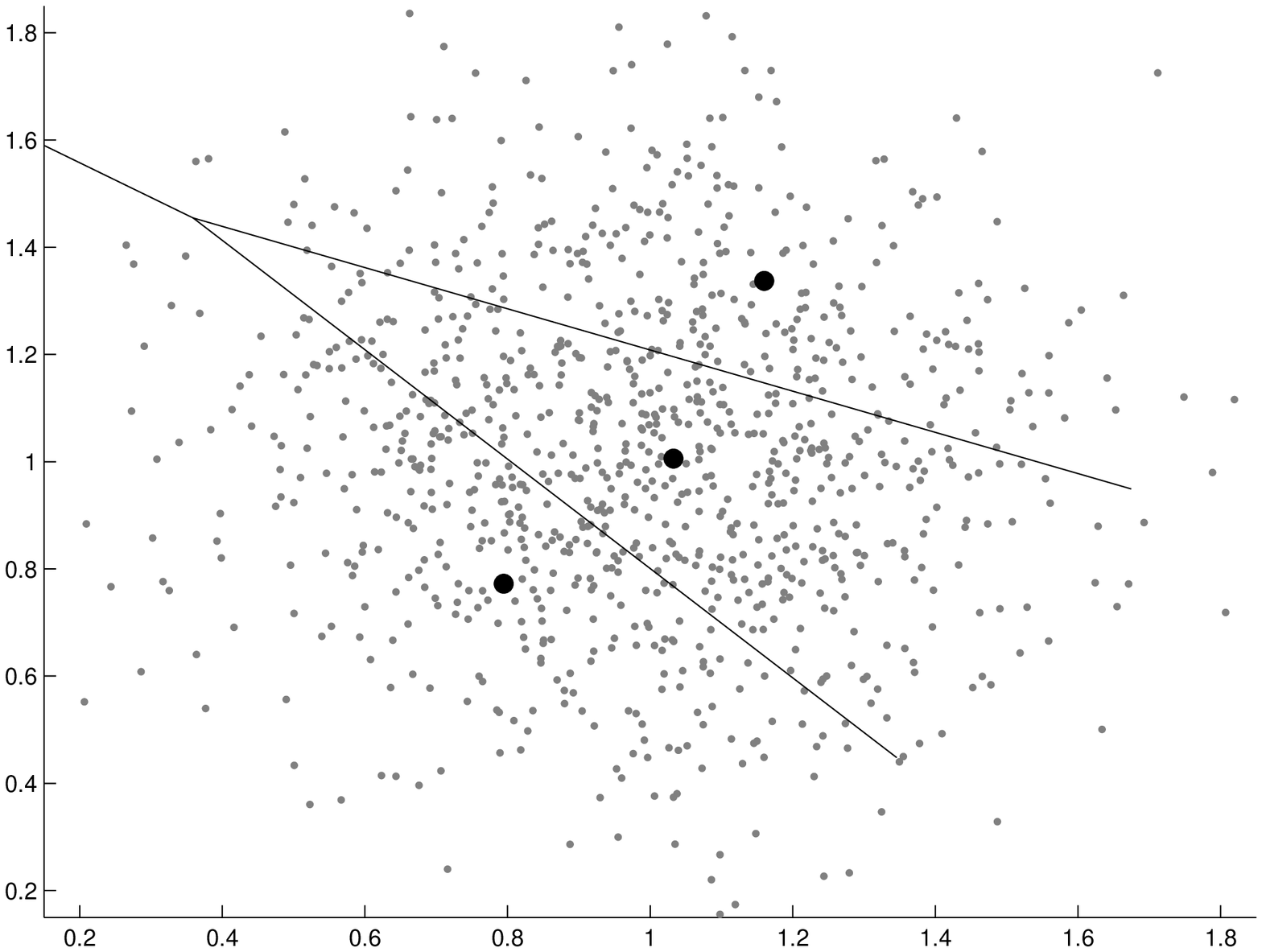}
\caption{Level 1}
\label{fig:ktlev1}
\end{figure}

\begin{figure}
\centering
\includegraphics[width=75mm]{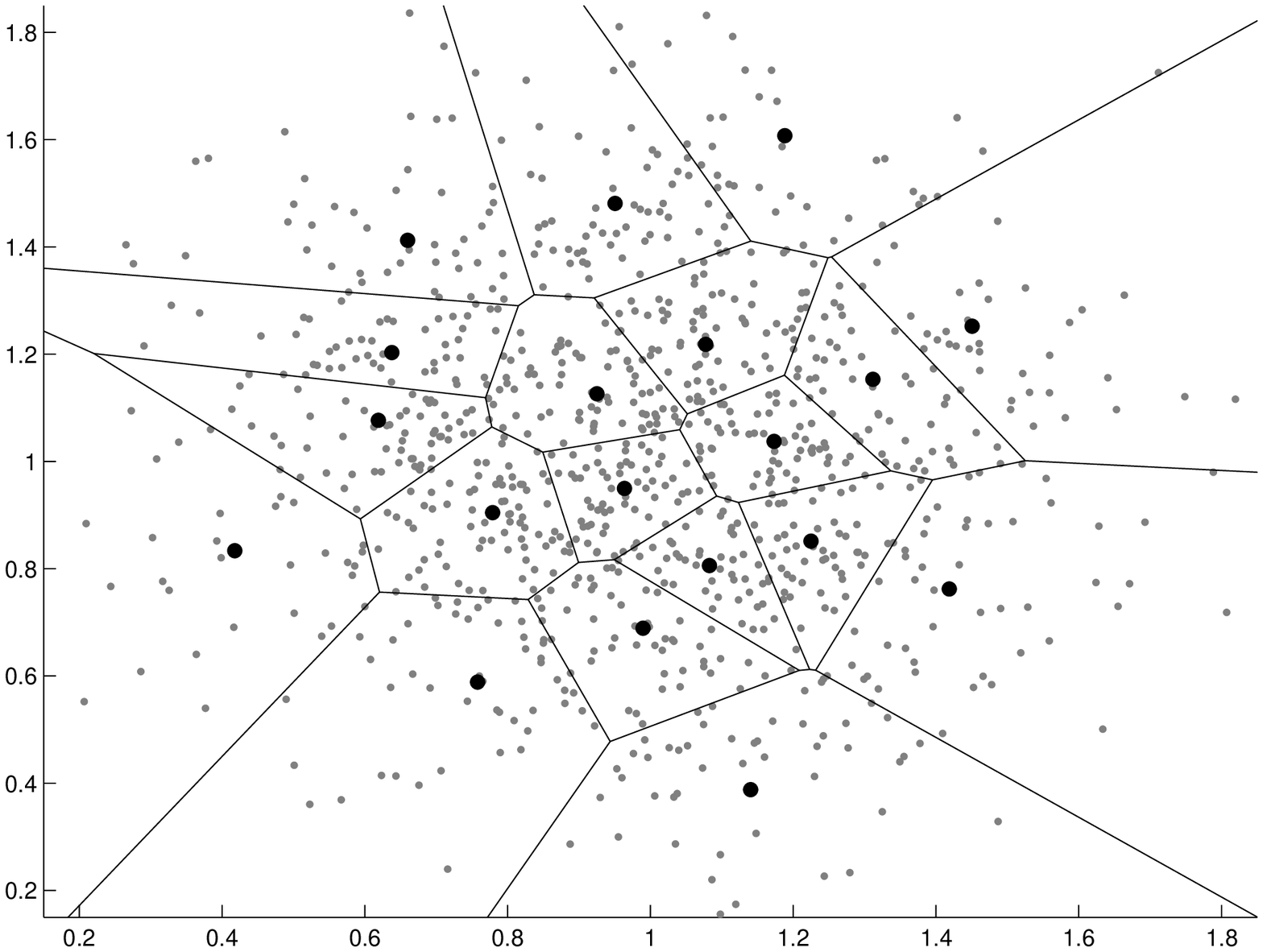}
\caption{Level 2}
\label{fig:ktlev2}
\end{figure}

\begin{figure}
\centering
\includegraphics[width=75mm]{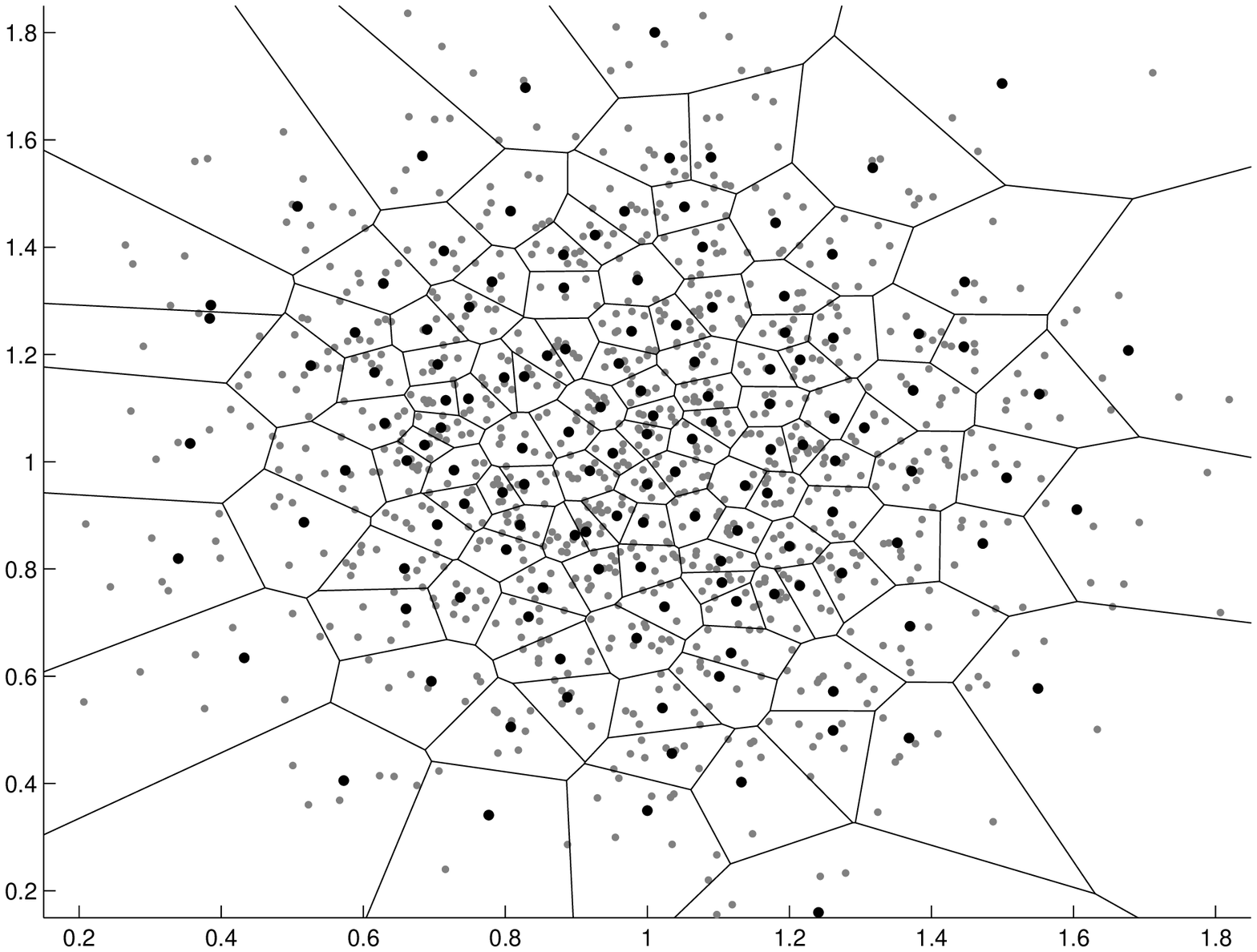}
\caption{Level 3}
\label{fig:ktlev3}
\end{figure}

Figures \ref{fig:ktlev1} to \ref{fig:ktlev3} are K-tree clusters in two dimensions. 1000 points were drawn from a random normal distribution with a mean of 1.0 and standard deviation of 0.3. The order of the K-tree, $m$, was 11. The grey dots represent the data set, the black dots represent the centroids and the lines represent the Voronoi tessellation of the centroids. Each of the data points contained within each tile of the tessellation are the nearest neighbours of the centroid and belong to the same cluster. It can be seen that the probability distribution is modelled at different granularities. The top level of the tree is level 1. It is the coarsest grained clustering. In this example it splits the distribution in three. Level 2 is more granular and splits the collection into 19 sub-clusters.  The individual clusters in level 2 can only be arrived at through a nearest neighbour association with a parent cluster in level 1 of the tree.  Level 3 is the deepest level in the tree consisting of cluster centroids. The 4th level is the data set of vectors that were inserted into the tree.

\subsection{Building K-tree}

\begin{figure*}
\centering
\includegraphics[scale=0.13]{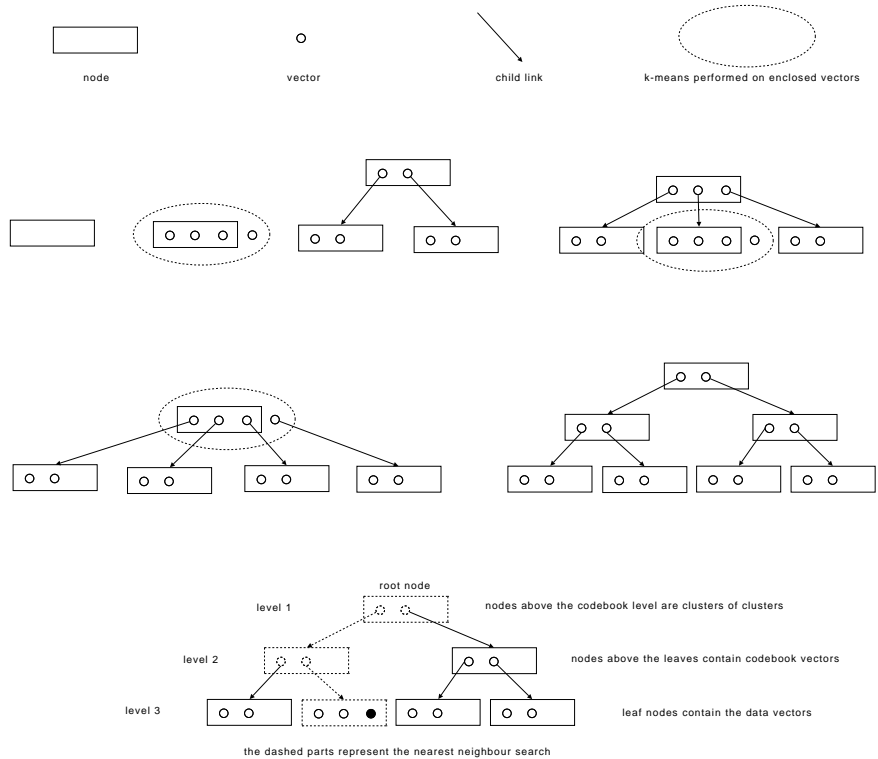}
\caption{K-tree Construction}
\label{fig:ktree2}
\end{figure*}

The K-tree is constructed dynamically as data vectors arrive. Initially the tree contains a single empty root node at the leaf level. Vectors are inserted via a nearest neighbour search, terminating at the leaf level. The root of an empty tree is a leaf, so the first $m$ data vectors are stored in the root, at which point the node becomes full. When the $m + 1$ vector arrives the root is split using k-means where $k = 2$, clustering all $m + 1$ vectors into two clusters. The two centroids that result from k-means are then promoted to become the centroids in a new root. The vectors associated with each centroid are placed into a child node. This promotion process has created a new root and two leaf nodes in the tree. The tree is now two levels deep. Insertion of a new data vector follows a nearest neighbour search to find the closest centroid in the root. The vector is inserted into the associated child. When a new vector is inserted the centroids are updated recursively along the nearest neighbour search path, all the way back to the root node. The propagated means are weighted by the number of data vectors contained beneath them. This ensures that any centroid in K-tree is the mean vector of all the data vectors contained in the associated sub tree. This insertion process continues, splitting leaves when they become full, until the root node itself becomes full. K-means is then run on the root node containing centroids. The vectors in the new root node become centroids of centroids.  As the tree grows, internal and leaf nodes are split in the same manner. The process of promotion can potentially propagate to cause a full root node at which point the construction of a new root follows and the tree depth is increased by one.  At all times the tree is guaranteed to be height balanced. Although the tree is always height balanced nodes can contain as little as one vector. In this case the tree will contain many more levels than a tree where each node is half full. Section \ref{fig:ktree2} shows this construction process for a K-tree of order three $(m = 3)$.

\subsection{Sparsity and K-tree}

K-tree was originally designed to operate with dense vectors. When a sparse representation is used performance degrades even though there is significantly less data to process. The clusters in the top levels of the tree are means of most of the terms in the collection and are not sparse at all. The algorithm updates cluster centres along the insertion path in the tree. Since document vectors have very high dimensionality this becomes a very expensive process. 

The medoid K-tree \cite{DeVries2009} extended the algorithm to use a sparse representation and replace centroids with document examples. This improved run-time performance and decreased memory usage. Unfortunately it decreased quality when using sparse document vectors. The document examples in the root of the tree were almost orthogonal to new documents being inserted. The documents were unlikely to have meaningful overlap in vocabulary.

The approach taken by De Vries and Geva at INEX 2008 \cite{DeVries2008} is a simple approach to dimensionality reduction or feature selection. It is called TF-IDF culling and it is performed by ranking terms. A rank is calculated by summing all weights for each term. The weights are the BM25 weight for each term in each document. This can also be explained as the sum of the column vector in the document by term matrix. The top n terms with the highest rank are selected, where n is the desired dimensionality. This works particularly well with term occurrences due to the Zipf law distribution of terms \cite{Zipf1949}. The collection frequency of a term is inversely proportional to its rank according to collection frequency. Most of the term weights are contained in the most frequent terms.

\section{Random Indexing}
\label{sec:ri}

Random Indexing (RI) \cite{Sahlgren2005} is an efficient, scalable and incremental approach to the word space model. Word space models use the distribution of terms to create high dimensional document vectors. The directions of these document vectors represent various semantic meanings and contexts.

Latent Semantic Analysis (LSA) \cite{Deerwester1990} is a popular word space model. LSA creates context vectors from a document term occurrence matrix by performing Singular Value Decomposition (SVD).  Dimensionality reduction is achieved through projection of the document term occurrence vectors onto the subspace spanned by the vectors with the largest Eigen values in the decomposition. This projection is optimal in the sense that it minimises the variance between the original matrix and the projected matrix. In contrast, Random Indexing first creates random context vectors of lower dimensionality, and then combines them to create a term occurrence matrix in the dimensionally reduced space.  Each term in the collection is assigned a random vector, and the document term occurrence vector is then a superposition of all the term random vectors. There is no matrix decomposition and hence the process is efficient.  

The RI process is conceptually very different from LSA and does not have the same optimality properties. The context vectors used by RI should optimally be orthogonal. Nearly orthogonal vectors can be used and have been found to perform similarly \cite{Bingham2001}. These vectors can be drawn from a random Gaussian distribution. The Johnson and Linden-Strauss lemma \cite{Johnson1984} states that if points are projected into a randomly selected subspace of sufficiently high dimensionality, then the distances between the points are approximately preserved. The same topology that exists in the higher dimensional space is reflected in the lower dimensional randomly selected subspace. Consequently, RI offers low complexity dimensionality reduction while still preserving the topological relationships amongst document vectors.

\subsection{Random Indexing Definition}

In RI, each dimension in the original space is given a randomly generated index vector. The index vectors are high dimensional, sparse and ternary. Sparsity is controlled via a seed length that specifies the number of randomly selected non-zero dimensions. Ternary vectors consist of randomly distributed +1 and -1 values in the non-zero dimensions.

In the context of document clustering, RI can be viewed as a matrix multiplication of a document by term matrix $D$ and a term by index-vector matrix $I$. Alternatively, $I$ can be referred to as a random projection matrix. Each row vector in $D$ represents a document, each row vector in $I$ is an index vector, $n$ is the number of documents, $t$ is the number of terms and $r$ is the dimensionality of the reduced spaced. $R$ is the reduced matrix where each row vector represents a document.

\begin{equation}\label{eq:ri}
D_{n \times t}I_{t \times r} = R_{n \times r}
\end{equation}

RI has several advantages. It can be performed incrementally and on-line as data arrives. Any document can be indexed (i.e. encoded as an RI vector) independently from all other documents in the collection.  This eliminates the need to build and store the entire document by term matrix. Additionally, newly encountered dimensions (terms) in the document collection are easily accommodated without having to recalculate the projection of previously encoded documents. In contrast, SVD requires global analysis where the number of documents and terms are fixed. The time complexity of RI is also very attractive. It is linear in the number of terms in a document and independent of collection size. 

\subsection{Choice of Index Vectors}

The index vectors used in RI were chosen to be sparse and ternary. Ternary index vectors for RI were introduced by Achlioptas \cite{Achlioptas2003} as being well suited for database environments. The primary concern of sparse index vectors is reducing time and space complexity. Bingham and Mannila \cite{Bingham2001} run experiments indicating that sparse index vectors do not affect the quality of results. This is not the only choice when creating index vectors. Kanerva \cite{Kanerva1994} introduces binary spatter codes. Plate \cite{Plate1994} explores Holographic Reduced Representations that consist of dense vectors with floating point values.

\subsection{Random Indexing Example}

In practice, to construct a document vector, the document vector is initially set to zero, and then the sparse index vector for each term in the document is added to the document vector. The weight of the added term index vector may be determined by TF-IDF or another weighting scheme. When all terms have been added, the document vector is normalised to unit length. There is no need to explicitly form the random projection matrix in Equation \ref{eq:ri} up-front. The random index vectors for each term can be generated and stored as they are first encountered. The fact that each index vector is sparse means that the vectors use less memory to store and are faster to add.

The effect of this approach is that each document will have a particular signature that can be compared with other documents via cosine similarity. The document signature is thus a vector on the unit hyper-sphere.

\begin{figure}
\centering
\includegraphics[width=75mm]{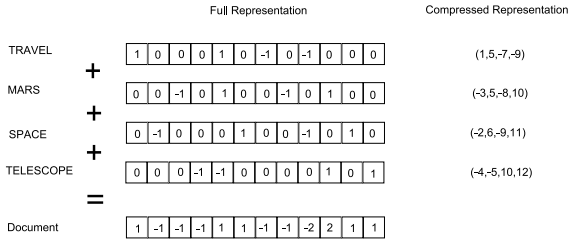}
\caption{Random Indexing Example}
\label{fig:ri_example}
\end{figure}

In the simple scenario in Figure \ref{fig:ri_example} the index vectors for the four words travel, mars, space and telescope, are added to the document vector as they are encountered in the text of the document. Afterwards, the document should be normalised.

The sparse index vectors can be efficiently stored by simply storing the position of the non-zero entries with the sign of the position indicating whether it is one or negative one.

\subsection{Random Indexing K-tree}

The time complexity of K-tree depends on the length of the document vectors. K-tree insertion incurs two costs, finding the appropriate leaf node for insertion and k-means invocation during node splits. It is therefore desirable to operate with lower dimensional vector representation.   

The combination of RI with K-tree is a good fit. Both algorithms operate in an on-line and incremental mode. This allows it to track the distribution of data as it arrives and changes over time. K-tree insertions and deletions allow flexibility when tracking data in volatile and changing collections. Furthermore, K-tree performs best with dense vectors, such as those produced by RI.

\section{Document Representation}
\label{sec:docrep}

The INEX 2008 XML Mining collection was used to complete the experiments. It contains 114,366 documents that are a subset of the XML Wikipedia corpus \cite{Denoyer2006}. 15 different categories were provided for the documents.

Document content was represented with BM25 \cite{Robertson1997}. Stop words were removed and the remaining terms were stemmed using the Porter algorithm \cite{Porter2006}. BM25 is determined by term distributions within each document and the entire collection. BM25 works with similar concepts as TF-IDF except that is has two tuning parameters. The BM25 tuning parameters were set to the same values as used for TREC \cite{Robertson1997}, $K1 = 2$ and $b = 0.75$. $K1$ influences the effect of term frequency and $b$ influences document length.

Links were represented using LF-IDF \cite{DeVries2008}. This resulted in a document-to-document link matrix. If there is a link between documents $i$ and $j$ then a value of one is added to position $i, j$ and $j, i$ in the matrix. If two documents both link to each other a value of two is recorded in their respective vectors. Each row vector of the matrix represents a document as a vector of link frequencies to and from other documents.

The motivation behind this representation is that documents with similar content will link to similar documents. For example, in the current Wikipedia both car manufacturers BMW and Jaguar link to the Automotive Industry document. Link frequencies were weighted with the same Inverse Document Frequency heuristic from TF-IDF. The idea is to decrease the weight of highly frequent links and increase the weight of less frequent links. Links to year documents in the Wikipedia are examples of ``stop links'' that are weighted down by this heuristic. Unlike term frequencies in TF-IDF the link frequencies in LF-IDF are not normalised. De Vries and Geva \cite{DeVries2008} found that normalising link frequencies decreased classification performance.

When document and link representations are combined they are both converted to unit vectors and concatenated. Converting each representation to unit vectors ensures that the weights of one representation do not dominate the other. De Vries and Geva \cite{DeVries2008} found this to be effective for classification.

\section{Experimental Setup}
\label{sec:setup}

Experiments have been run to measure the quality difference between various configurations of K-tree. Section \ref{sec:modifications} describes the modifications made to K-tree. Table \ref{table:ktree_configs} lists all the configurations tested.

The following conditions were used when running the experiments.
\begin{enumerate}
\item Each K-tree configuration was run a total of 20 times.
\item The documents were inserted in a different random order each time K-tree is built. 
\item If RI was used, the index vectors were generated statistically independently each time K-tree was built.
\item For each K-tree built, k-means++ \cite{Arthur2007} was run 20 times on the codebook vectors to create 15 clusters.
\item All document vectors were unitised after performing dimensionality reduction.
\end{enumerate}

The conditions listed above resulted in 400 measurements for each K-tree configuration. For each of the 20 K-trees built, k-means++ was run 20 times. The repetition of the experiments is to measure the variance caused by the random insertion order into K-tree, the randomised seeding process in k-means in the modified K-tree and the randomised seeding process of k-means++.

Assessment of clustering quality is based on the INEX XML Mining track.  The set of 114,366 documents, belonging to 15 classes were used to evaluate clustering quality of INEX submissions. The cluster labels are taken from the Wikipedia itself. K-tree generates clusters in an unsupervised manner, and it is not necessarily going to produce 15 clusters at a particular level in the tree.  In order to re-use the INEX test collection, it was necessary to post process the K-tree and to reduce a cluster level in the tree to 15 clusters by using k-means++. Note that this is a low cost operation involving only a small number of vectors, which is not required in an ordinary application. It is done for the sole purpose of producing comparable results with the INEX benchmark data. The same approach was taken at INEX 2008 by De Vries and Geva \cite{DeVries2008}. For a comparison of entropy and purity to be meaningful they have to be measured on the same number of clusters.

Micro averaged purity and entropy are compared. Micro averaging weights the score of a cluster by its size. Purity and entropy are calculated by comparing the clustering solution to the labels provided. A higher purity score indicates a higher quality solution because the clusters are more pure with respect to the ground truth. A lower entropy score indicates a higher quality solution because there is more order with respect to the ground truth.

\section{Experimental Results}
\label{sec:results}

\begin{figure}
\centering
\includegraphics[width=75mm]{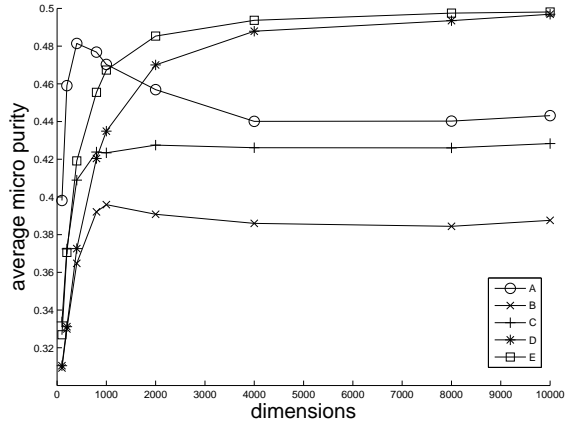}
\caption{Purity Versus Dimensions}
\label{fig:purity}
\end{figure}

\begin{figure}
\centering
\includegraphics[width=75mm]{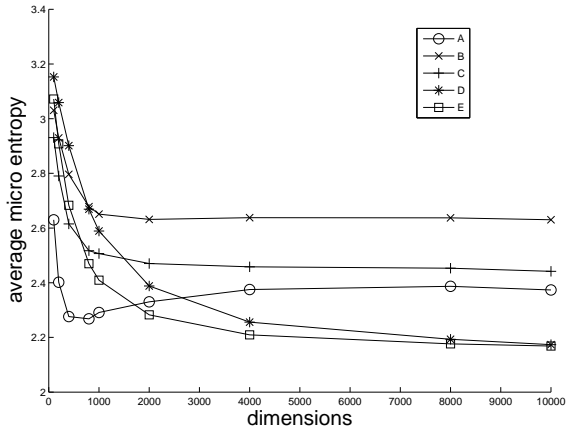}
\caption{Entropy Versus Dimensions}
\label{fig:entropy}
\end{figure}

Tables \ref{table:ktA} to \ref{table:ktE} contain results for the K-tree configurations tested listed in Table \ref{table:ktree_configs}. Table \ref{table:symbols} lists the meaning of the symbols used. Figures \ref{fig:purity} and \ref{fig:entropy} are graphical representations of the average micro purity and entropy.

The unmodified K-tree using TF-IDF culling and BM25 had unexpected results as seen in Table \ref{table:ktA}. The average micro purity and entropy peaked at 400 dimensions. Performing this dimensionality reduction at these lower dimensions had not been performed before. This is an interesting and unexpected result and future experiments will need to determine if the phenomenon occurs in different corpora.

Improvements in micro purity have been tested for significance via t-tests. The null hypothesis is that both results come from the same distribution with the same mean. In this case they are not significantly different. If the null hypothesis is rejected then the difference is statistically significant.

The modifications made to K-tree for use with RI had a significant impact. The unmodified K-tree and modified K-tree were compared. Specifically, configurations B and D, and configurations C and E were tested against each other. All dimensions were compared against each other. The improved performance of the modified K-tree was statistically significant for all dimensions (100 vs 100, 200 vs 200 and so on) with a p-value of 0 or extremely close to 0 ($p < 1 \times 10^{-100}$).

The modified K-tree using RI was tested with two representations. Configurations D and E were tested at all dimensions. The null hypothesis was rejected at all dimensions except 10000. This means that BM25 performed significantly better than the BM25 + LF-IDF representation at all dimensions except 10000. At 10000 dimensions the difference was not considered statistically significant with a p-value of 0.3. The increased performance of this representation in classification did not apply to clustering when using RI. The LF-IDF representation may be interfering with the BM25 representation and approaches such as reducing the weight of LF-IDF in the RI process or performing RI separately on each representation and then concatenating the reduced vectors may improve performance. Running k-means on the full sparse vectors will also indicate if RI is responsible for this. Further experimentation is required to provide more evidence for this result.

The unexpected results in configuration A were tested against the best RI configuration, E. The highest average at 400 dimensions in configuration A was tested against all dimensions in configuration E (400 vs 100, 400 vs 200, 400 vs 400, 400 vs 1000 and so on). The RI K-tree, configuration E, became statistically more significant at 2000 dimensions with a p-value of $1.48 \times 10^{-6}$ and thus rejected the null hypothesis. For dimensions 4000 through 10000, the performance difference was statistically significant, with a p-value of 0 in all cases. Thus, RI K-tree improves results, even over the unexpected high results of configuration A, by embedding the original 200,000 dimensional term space into at least a 2000 dimension reduced space.

\subsection{INEX Results}

The INEX XML Mining track is a collaborative evaluation forum where research teams improve approaches in supervised and unsupervised machine learning with XML documents. Participants make submissions and the evaluation results are later released.

The RI K-tree in configuration E performs on average at a comparable level to the best results submitted to the INEX 2008 XML Mining track. The top two results from the track had a micro purity of 0.49 and 0.50. These are not average scores for the approaches but the best results participants found. The RI K-tree in configuration E had a maximum micro entropy of 0.55. This is 10\% greater than the INEX submissions.

\begin{table}
\begin{center}
\begin{tabular}{ccc}
\hline\noalign{\smallskip}
ID & K-tree & Representation \\
\noalign{\smallskip}
\hline
\noalign{\smallskip}
A & Unmodified & TF-IDF Culling, BM25 \\
B & Unmodified & RI, BM25 + LF-IDF \\
C & Unmodified & RI, BM25 \\
D & Modified & RI, BM25 + LF-IDF \\
E & Modified & RI, BM25 \\
\hline
\end{tabular}
\caption{K-tree Test Configurations}
\label{table:ktree_configs}
\end{center}
\end{table}

\begin{table}
\begin{center}
\begin{tabular}{cc}
\hline\noalign{\smallskip}
Symbol & Meaning \\
\noalign{\smallskip}
\hline
\noalign{\smallskip}
$\alpha$ & Average Micro Entropy \\
$\beta$ & Standard Deviation of $\alpha$ \\
$\gamma$ & Average Micro Purity \\
$\delta$ & Standard Deviation of $\gamma$ \\
\hline
\end{tabular}
\caption{Symbols for Results}
\label{table:symbols}
\end{center}
\end{table}

\begin{table}
\begin{center}
\begin{tabular}{ccccc}
\hline\noalign{\smallskip}
Dimensions & $\alpha$ & $\beta$ & $\gamma$ & $\delta$ \\
\noalign{\smallskip}
\hline
\noalign{\smallskip}
100 & 2.6299 & 0.0194 & 0.3981 & 0.0067 \\
200 & 2.4018 & 0.0207 & 0.4590 & 0.0085 \\
400 & 2.2762 & 0.0263 & 0.4814 & 0.0093 \\
800 & 2.2680 & 0.0481 & 0.4768 & 0.0155 \\
1000 & 2.2911 & 0.0600 & 0.4703 & 0.0192 \\
2000 & 2.3302 & 0.0821 & 0.4569 & 0.0254 \\
4000 & 2.3751 & 0.1103 & 0.4401 & 0.0331 \\
8000 & 2.3868 & 0.1068 & 0.4402 & 0.0300 \\
10000 & 2.3735 & 0.1062 & 0.4431 & 0.0306 \\
\hline
\end{tabular}
\caption{A: Unmodified K-tree, TF-IDF Culling, BM25}
\label{table:ktA}
\end{center}
\end{table}

\begin{table}
\begin{center}
\begin{tabular}{ccccc}
\hline\noalign{\smallskip}
Dimensions & $\alpha$ & $\beta$ & $\gamma$ & $\delta$ \\
\noalign{\smallskip}
\hline
\noalign{\smallskip}
100 & 3.0307 & 0.0149 & 0.3093 & 0.0045 \\
200 & 2.9295 & 0.0206 & 0.3300 & 0.0079 \\
400 & 2.7962 & 0.0379 & 0.3648 & 0.0143 \\
800 & 2.6781 & 0.0718 & 0.3921 & 0.0236 \\
1000 & 2.6509 & 0.0842 & 0.3959 & 0.0260 \\
2000 & 2.6315 & 0.1262 & 0.3908 & 0.0345 \\
4000 & 2.6380 & 0.1451 & 0.3860 & 0.0356 \\
8000 & 2.6371 & 0.1571 & 0.3844 & 0.0382 \\
10000 & 2.6302 & 0.1540 & 0.3876 & 0.0385 \\
\hline
\end{tabular}
\caption{B: Unmodified K-tree, Random Indexing, BM25 + LF-IDF}
\label{table:ktB}
\end{center}
\end{table}

\begin{table}
\begin{center}
\begin{tabular}{ccccc}
\hline\noalign{\smallskip}
Dimensions & $\alpha$ & $\beta$ & $\gamma$ & $\delta$ \\
\noalign{\smallskip}
\hline
\noalign{\smallskip}
100 & 2.9308 & 0.0213 & 0.3337 & 0.0089 \\
200 & 2.7902 & 0.0335 & 0.3724 & 0.0126 \\
400 & 2.6151 & 0.0417 & 0.4089 & 0.0116 \\
800 & 2.5170 & 0.0703 & 0.4238 & 0.0197 \\
1000 & 2.5066 & 0.0858 & 0.4234 & 0.0240 \\
2000 & 2.4701 & 0.0938 & 0.4275 & 0.0258 \\
4000 & 2.4581 & 0.0979 & 0.4261 & 0.0271 \\
8000 & 2.4530 & 0.1139 & 0.4260 & 0.0318 \\
10000 & 2.4417 & 0.1019 & 0.4283 & 0.0283 \\
\hline
\end{tabular}
\caption{C: Unmodified K-tree, Random Indexing, BM25}
\label{table:ktC}
\end{center}
\end{table}

\begin{table}
\begin{center}
\begin{tabular}{ccccc}
\hline\noalign{\smallskip}
Dimensions & $\alpha$ & $\beta$ & $\gamma$ & $\delta$ \\
\noalign{\smallskip}
\hline
\noalign{\smallskip}
100 & 3.1527 & 0.0227 & 0.3105 & 0.0047 \\
200 & 3.0589 & 0.0266 & 0.3312 & 0.0065 \\
400 & 2.9014 & 0.0259 & 0.3726 & 0.0065 \\
800 & 2.6690 & 0.0336 & 0.4204 & 0.0085 \\
1000 & 2.5890 & 0.0319 & 0.4349 & 0.0090 \\
2000 & 2.3882 & 0.0428 & 0.4700 & 0.0129 \\
4000 & 2.2558 & 0.0443 & 0.4879 & 0.0144 \\
8000 & 2.1933 & 0.0473 & 0.4935 & 0.0162 \\
10000 & 2.1735 & 0.0496 & 0.4969 & 0.0171 \\
\hline
\end{tabular}
\caption{D: Modified K-tree, Random Indexing, BM25 + LF-IDF}
\label{table:ktD}
\end{center}
\end{table}

\begin{table}
\begin{center}
\begin{tabular}{ccccc}
\hline\noalign{\smallskip}
Dimensions & $\alpha$ & $\beta$ & $\gamma$ & $\delta$ \\
\noalign{\smallskip}
\hline
\noalign{\smallskip}
100 & 3.0717 & 0.0263 & 0.3269 & 0.0074 \\
200 & 2.9078 & 0.0291 & 0.3706 & 0.0087 \\
400 & 2.6832 & 0.0293 & 0.4191 & 0.0077 \\
800 & 2.4696 & 0.0350 & 0.4555 & 0.0106 \\
1000 & 2.4093 & 0.0399 & 0.4673 & 0.0115 \\
2000 & 2.2826 & 0.0422 & 0.4853 & 0.0137 \\
4000 & 2.2094 & 0.0416 & 0.4937 & 0.0141 \\
8000 & 2.1764 & 0.0429 & 0.4975 & 0.0149 \\
10000 & 2.1686 & 0.0440 & 0.4981 & 0.0161 \\
\hline
\end{tabular}
\caption{E: Modified K-tree, Random Indexing, BM25}
\label{table:ktE}
\end{center}
\end{table}

\section{Conclusion}
\label{sec:conclusion}

RI K-tree was introduced as an attractive approach for large scale document clustering. This is the first time RI and K-tree have been combined. The results show that RI K-tree improves quality of clustering results, even over the unexpected results found when using TF-IDF culling. Further experiments are required to determine if the unexpected effect of TF-IDF culling at low dimensions is an anomaly or actually exists in many collections. Additionally, RI K-tree is an efficient and high quality approach to overcome previous problems with sparse representations when using K-tree. Unfortunately the combination of BM25 and LF-IDF representations did not improve results in clustering as they did in earlier classification results.

\begin{small}
\bibliography{adcs09}
\end{small}

\end{document}